\documentclass[journal]{IEEEtran}
\usepackage{graphicx}
\usepackage[font={small}]{caption}	
\usepackage[font=small,labelfont={small}]{caption}
\usepackage{amsmath}
\usepackage{siunitx}
\usepackage{subcaption}
\begin{document}

\title{\underline{D}istance \underline{A}ware \underline{R}elaying \underline{E}nergy-efficient:\\
 DARE to Monitor Patients in Multi-hop\\ Body Area Sensor Networks}

\author{A. Tauqir$^{1}$, N. Javaid$^{2,3}$, S. Akram$^{1}$, A. Rao$^{2}$, S. N. Mohammad$^{2}$\\\vspace{0.4cm}
$^{1}$Institute of Space Technology, Islamabad, Pakistan.\\
$^{2}$Dept. of Electrical Engineering, COMSATS Institute of IT, Islamabad, Pakistan.\\
$^{3}$CAST, COMSATS Institute of IT, Islamabad, Pakistan.}

\maketitle

\begin{abstract}
In recent years, interests in the applications of Wireless Body Area Sensor Network (WBASN) is noticeably developed. WBASN is playing a significant role to get the real time and precise data with reduced level of energy consumption. It comprises of tiny, lightweight and energy restricted sensors, placed in/on the human body, to monitor any ambiguity in body organs and measure various biomedical parameters. In this study, a protocol named Distance Aware Relaying Energy-efficient (DARE) to monitor patients in multi-hop Body Area Sensor Networks (BASNs) is proposed. The protocol operates by investigating the ward of a hospital comprising of eight patients, under different topologies by positioning the sink at different locations or making it static or mobile. Seven sensors are attached to each patient, measuring different parameters of Electrocardiogram (ECG), pulse rate, heart rate, temperature level, glucose level, toxins level and motion. To reduce the energy consumption, these sensors communicate with the sink via an on-body relay, affixed on the chest of each patient. The body relay possesses higher energy resources as compared to the body sensors as, they perform aggregation and relaying of data to the sink node. A comparison is also conducted conducted with another protocol of BAN named, Mobility-supporting Adaptive Threshold-based Thermal-aware Energy-efficient Multi-hop ProTocol (M-ATTEMPT). The simulation results show that, the proposed protocol achieves increased network lifetime and efficiently reduces the energy consumption, in relative to M-ATTEMPT protocol.
\end{abstract}


\section{Introduction}
\IEEEPARstart{T}{he} field of Computer Science is constantly emerging with an aim to process larger data sets and maintain higher levels of connectivity. At the same time, advancements in miniaturization allow for increased mobility and ease of access.
Wireless Sensor Network (WSN) is emerging as a promising technology for a wide variety of applications in the fields of remote health monitoring, home/health care, medicine, multimedia, sports and many others as shown in Fig. 1. All these can be converged in one of the sub-classes of WSN, named as WBASN.

\begin{figure}[htbp]
  \begin{center}
  \includegraphics[scale=0.4]{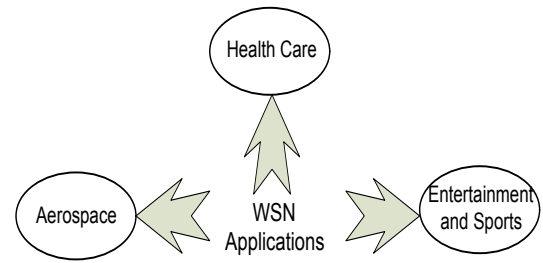}
  \end{center}
  \begin{center}
\caption {WSN Applications}
\end{center}
\end{figure}

WBASN consist of tiny and lightweight sensors which, have limited energy and processing resources. These sensors can either be in-vivo sensors i.e. implanted inside the body or can be wearable sensors [1] attached on the body, depending upon the application requirements.

In the field of sports, WBASN can be deployed by using a heart rate measuring sensor placed on players body to replace him/her when, the rate reaches a critical level while playing, in order  to avoid the bad performance of a particular team. In Aerospace, WBASN can be beneficial where, a flying system can extract the information from multiple sensors, warn the pilot in case of emergencies, and provide feedback during supervised recovery or normal activity \cite{tufail2009wearable}.

In the medical field, WBASN makes use of the tiny sensors for detecting and monitoring different biological characteristics of a human body. Both the aspects of connectivity as well as miniaturization are covered in WBASN. In terms of patient monitoring, a patient at home or in hospital can be equipped with different physical parameters measuring sensors. The monitored signals are then collected or aggregated by a personal device, e.g. Personal Digital Assistant (PDA) which, acts as a relay node and transmits them to health-care professional for health monitoring or to some external network acting as a sink. The sink node acts as a bridge between the hospital’s (wired) network, and the BAN.

The advantage of this entire topology is that the patient is not required to make frequent visits to the hospital, thereby, reducing transportation costs and fear of any miss checkup visits which, might be occurring due to busy life routines and schedules. And also, when in hospital, no need for a patient to stay continuously on bed or can move freely. This whole process, provides real-time feedback and sometimes faster diagnosis and enables the patient to receive immediate medication or medical assistant.

In this paper, a relaying protocol named, DARE to monitor patients in multi-hop BASNs is proposed. In DARE, different hospital ward scenarios are undertaken. The ward consists of eight patients arranged on beds, properly aligned and equipped with seven different sensors monitoring vital signs and one body relay each. Out of these seven sensors, two of the sensors monitor the patient only, when there reaches a threshold level. Thus, the protocol also supports the monitoring of multiple data types. In each scenario, the placement of sensors is kept fixed while, the sink node is either made mobile or static. To verify the accuracy of the protocol, it is also compared with another BAN protocol called M-ATTEMPT. Simulations are carried out in MATLAB. The comparison results exhibit improvements in the proposed protocol in terms of reduced energy consumption by the biomedical sensor nodes as well as an elongated network lifetime.

The remainder of the paper is organized as follows. Section II gives an overview on the previous research works on BASN communications. In section III, motivation for proposing new protocol is mentioned. Proposed technique is explained in section IV. Section V compares the plots between the proposed and the compared protocol. Finally, section VI concludes the paper.

\section{Related Work}

In terms of network organization, routing protocols can be classified into three categories as flat-based protocols, hierarchical-based and location-based protocols. Restricting the discussion only to first two categories, flat-based protocols comprise of the nodes that are homogeneous in terms of their energy resources and sensing capabilities. While, hierarchical-based protocols incorporate sensors that have different responsibilities. Some of the sensors are superior in terms of energy and thus, are overloaded in relative to others. These nodes act as relay nodes which, transmit and aggregate the received data from energy-constrained nodes and ultimately send it to a destination node or sink.

Similarly, the routing protocols can also be categorized into single-hop or multi-hop communication protocols. In WSN, single-hop communication may be required in the cases where, central management is considered important. However, in BAN, it provides poor results as, the energy-constrained sensor nodes are over-burdened due to self transmitting data to a far-distant sink node which, exhaust the sensors earlier. In comparison to single-hop, multi-hop communication is considered favorable and applicable for BANs to protect the human body by excessive heat radiations, dissipated from the batteries of the sensor nodes. In multi-hop networks, in-vivo or on-body sensors transmit information to the nearby located relay node(s) which, then transmits data to external network for monitoring purpose. Hence, reduces the distance travelled by the data and also saves energy consumption of the sensor node.

There are different parameters e.g. network lifetime, energy consumption, propagation delay, network throughput, supporting multiple data types etc. upon which, wide researches are conducted to stabilize the BAN network. Very few works appear in the literature with the purpose of increasing network lifetime and improving energy efficiency of WBASN, using additional devices called relay nodes. By means of multi-hop communication, some of the parameters are achieved with good precision.

In \cite{reusens2009characterization}, E. Reusens $et$ $al.$  focus on increasing network lifetime by means of relaying and cooperation. First, the relay nodes perform relaying of traffic only so that, high amount of energy is available for communication purposes. Next, the relays cooperate in forwarding the data from one node towards the central device.

In \cite{ehyaie2009using}, A. Ehyaie $et$ $al.$ set an upper bound to determine the number of relay nodes, sensors and their respective distances to the sink. A relay network is defined as a network of relay nodes distributed along the body in combination with sensor nodes to serve as a transport network for the BSN sensor nodes. Hence, no sensor node needs to relay other sensor's traffic, any more. Each sensor node does single-hop communication while, relay nodes are responsible for multi-hop communication to the sink.

In \cite{elias2012energy}, Jocelyne Elias $et$ $al.$ propose an optimal design for WBANs, by studying the joint data routing and relay positioning problem in order to increase the network lifetime.

In \cite{braem2007need}, authors study propagation models for improving the energy efficiency. Using these models, calculations show that single-hop communication is inefficient, especially for the nodes far away from sink, however, multi-hop proves to be more efficient. In order to avoid hotspot links, extra nodes in the network, i.e. dedicated relay devices are introduced. The experimental results show that, these solutions significantly increase the network lifetime.

Authors in \cite{ain2012modeling}, derive analytical channel modeling and propagation characteristics of arm motion as spherical model. Four possible cases are presented where, transmitter and receiver lies inside or outside of the body.

In \cite{chen2010patient}, Baozhi Chen $et$ $al.$ introduce a new interference-aware WBAN, that can continuously monitor vital signs of multiple patients and efficiently prioritize data transmission based on patients condition.

In \cite{seo2010energy}, Su-Ho $et$ $al.$ suggest a heuristic adaptive routing algorithm  for  an  energy-efficient  configuration  management which,  can  reduce energy  consumption  while,  guaranteeing  QoS  for  the  emergency  data  in WBASNs. The  priority  and  vicinity  of  the  nodes  are  taken  into  account, for the selection of reachable parent nodes when,  the  nodes  are  disconnected  due  to  the  mobile  nature  of  human  body.

In \cite{watteyne2007anybody}, Thomas Watteyne $et$ $al.$ propose a protocol named AnyBody which, is a self-organization protocol comprising of sensor nodes, grouped into clusters. It focuses on relaying data via cluster heads to improve the routing and energy efficiency.

In \cite{javaid2013eddeec}, a novel clustering based routing technique for heterogeneous networks is proposed. It is based on changing the Cluster Head (CH) dynamically, according to some election probability. Simulation results show that, this proposed protocol achieves longer lifetime and stability period.

In \cite{otto2006wban}, Chris A. Otto $et$ $al.$ describe a prototype system for continual health monitoring at home. The system consists of an uninterrupted WBASN and a home health server. The WBASN sensors monitor user's heart rate and locomotive activity and periodically, upload time-stamped information to the home server. The home server may integrate this information into a local database for user's inspection or it may further forward the information to a medical server.

In \cite{javaid2013ubiquitous} and \cite{hayat2012energy}, path loss in WBASN and its impact on communication is presented with the help of simulations which, were performed for different models of In-Body communication and different factors (such as, attenuation, frequency, distance etc), influencing path loss in On-Body communications.

In \cite{wang2012distributed}, Changhong Wang  $et$ $al.$ propose a distributed WBASN for medical supervision. The system contains three layers: sensor network tier, mobile computing network tier and remote monitoring network tier. It provides collection, demonstration and storage of the vital information such as, ECG, blood oxygen, body temperature and respiration rate. The system demonstrates many advantages such as, low-power, easy configuration, convenient carrying, and real-time reliable data.

In \cite{abbasi2007survey}, A.A. Abbasi $et$ $al.$ suggest a scheme in which, nodes are grouped in a number of clusters. There is a special node within each cluster named as cluster head which, is responsible for collecting the data of other nodes in a cluster, pack them together and sent them to sink. Therefore, energy usage in cluster heads are much more than the other nodes to maximize the lifetime of information carrying sensors.

In \cite{javaid2013analyzing}, transmission delay of different paths through which, data is sent from sensor to health care center over heterogeneous multi-hop wireless channel is analyzed.

In \cite{javaid2013m}, N. Javaid $et$ $al.$ present an energy efficient routing algorithm for heterogeneous WBASNs. For on-demand data traffic, root node directly communicates with sink node and for normal data delivery, multi-hop communication is used. The proposed routing algorithm is thermal-aware which, senses the link Hot-spot and routes the data away from these links. The protocol focuses on reducing energy consumption and to maximize the number of packets, sent to network.

The performance of indoor localization schemes for optimal placement of wireless sensors in an area where, location tracking is required, is verified in \cite{rehman2013survey}. In this paper, the performance of Particle filtering and Kalman filtering based location tracking techniques using Bays algorithm and Hiden Markov Model (HMM) in terms of localization accuracy is presented.

\section{Motivation}
BASNs are gradually making advancements in the monitoring of humans in different fields like, players and athletes in a playing ground, astronauts in space, patients in the hospital or home and many others. In monitoring patients, BASN enables the medical personnel to provide an urgent medical aid in case of getting any threat or alarm. The sensors used in BASN consist of restricted energy resources. As, the sensors gather information about body organs and transmit the data ahead, much of the energy gets utilized due to which, excess heat is dissipated. As a result, the human body tissues get severely damaged. Keeping this issue in mind, many researches are conducted and still underway to reduce the energy consumption of the sensors and to enhance the lifetime of network so that, the patient need not to be operated very often to re-energize the depleted sensors.

In \cite{javaid2013m},the simulation results show that it’s processing creates very little propagation delay which is beneficial for the cases where, delays can not be tolerated. However, the nodes exhaust earlier in terms of rounds and also consume surplus energy.

To reduce energy utilization, enhance network lifetime and to improve the quality of living, this study presents a protocol, DARE. To verify the accuracy of the protocol, with respect to different network parameters it is compared with with a BAN protocol of M-ATTEMPT.

\section{Proposed Technique}
The proposed scheme is described in detail in below subsections.

\subsection{Network Topology}
The DARE protocol is based on inspecting a hospital ward with dimensions of “\SI{40 x 20}{ft^{2}}”, under five different scenarios in which, the patient’s different body organs are monitored to detect any ambiguity in it's normal functioning. The ward consists of eight beds where, each patient comprises of seven Body Sensors (BSs) placed on different positions and one Body Relay (BR), on the chest. The topology is kept same throughout the entire ward. This makes a total of fifty six BSs and eight BRs. Both Body Sensors and Body Relays have limited energy resources. The protocol assigns an energy of “\SI{0.3}{J}” to the BSs and “\SI{1}{J}” to the BR. A node of Main Sensor (MS) is also attached on bed in one of the scenarios to reduce the energy consumption. The MS is assumed to have either unlimited or at least very high energy resources as compared to other sensor types. The sink is considered to have unlimited energy resources which then, ultimately transmits the final form of information to the external network. The proposed protocol's patient is shown in Fig. 2.

\begin{figure}[htbp]
  \begin{center}
  \includegraphics[scale=0.7]{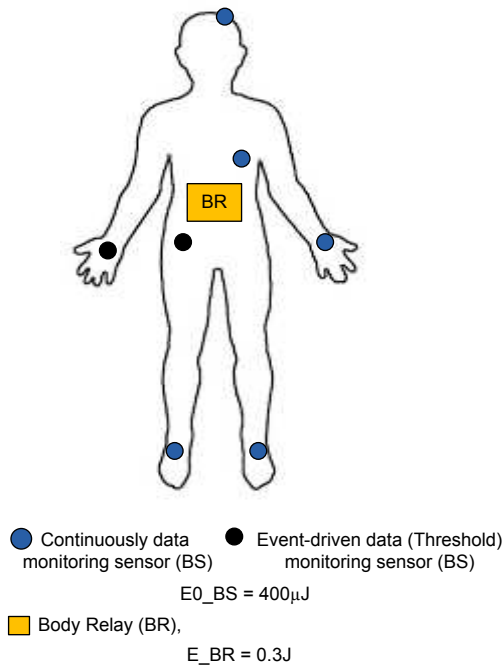}
  \end{center}
    \begin{center}
\caption {Scenario's Patient}
\end{center}
\end{figure}

The combination of Body Sensors and Body Relay makes this scheme a heterogeneous technique and let the information to travel in a multi-hop manner. The protocol assumes line of sight communication (LOS) between the sending and receiving nodes. The LOS co-efficient value is set to be 3.38 as, shown in TABLE I.

\begin{table}[htbp]
\centering
\setlength{\tabcolsep}{12pt}
   \begin{tabular}{| c | c | c | c |}
     \hline
     \textbf{Parameter} & \textbf{Value}  \\ \hline
     $E_{TXelec}$ & “\SI{16.7}{nJ/bit}“ \\ \hline
     $E_{RXelec}$ & “\SI{36.1}{nJ/bit}“ \\ \hline
     $E_{amp}(3.38)$ & “\SI{1.97}{nJ/bit}“ \\ \hline
     $E_{amp}(5.9)$ & “\SI{7.99}{nJ/bit}“ \\ \hline
   $w$ & “\SI{4000}{bits}“ \\ \hline
   \end{tabular}
    \caption{Parameters of energy model}
 \end{table}
\subsection{Types of data reporting}
The sensors measure ailment, either by continuously monitoring the data i.e. on the basis of time-driven events or whenever, a specific threshold level is reached i.e. on an event-driven basis, beneficial for pursuing critical monitoring. The sensors of glucose level and temperature level transmit the data whenever, the levels of both sensors either fall below the lower limit or exceed the higher limit, indicating an alarming condition. The values for low temperature value for a human body which can cause the patient, freeze to death is set to be “\SI{35}{^{o}C}” while, the high temperature level is set to be “\SI{40}{^{o}C}“ where, a patient absorbs more heat than can dissipate. The low and high critical levels for glucose are “\SI{110}{mg/dL}“ and “\SI{125}{mg/dL}“, respectively, showing the conditions of diabetes. The rest of the sensors continuously monitor the parameters of pulse rate and blood pressure, heart rate, ECG, motion and toxins level.

\subsection{Communication Flow}
The protocol monitors all the eight patients, one by one. In each patient, the BS conveys information to a corresponding BR in range, having higher energy resources which, then transmits the received information to the final destination node which, can either be a Main Sensor (MS) or sink, depending upon the particular scenario. Flow chart is given in Fig. 3.

\begin{figure}[h]
  \begin{center}
  \includegraphics[height=14cm,width=0.45\textwidth]{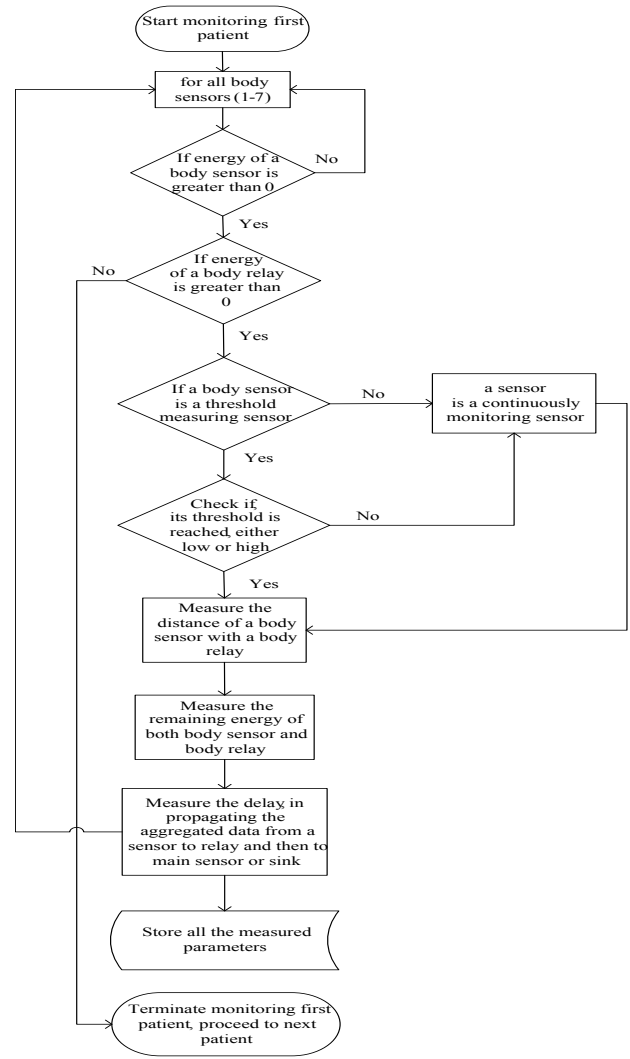}
  \end{center}
    \begin{center}
\caption {Flow chart for communication}
\end{center}
\end{figure}

The algorithm for monitoring one patient is as follows:
\begin{itemize}
  \item First, a BS is checked if, it is alive or not. If yes, then the algorithm checks for the BR to be alive. If found alive, then it checks whether the BS is a threshold measuring sensor or a continuously monitoring sensor.
  \item If the BS is a threshold measuring sensor, it checks if its low or high threshold level is reached. If yes, then it measures the distance between that BS and BR.
  \item Afterwards, it calculates the transmitted energy of BS and the received energy of BR.
  \item Then it estimates delay in propagating the data from BS to BR.
  \item If threshold is not reached, then the algorithm proceeds to check the other BSs in the same manner and then calculates the distance, remaining energy and delay. Finally, the estimated parameters are separately stored in different variables.
  \item This whole process continues till all the BSs are completely checked.
  \item After the BR receives all the sensors data, it then performs fusion process i.e. it aggregates the received information and transmit the data either directly to sink (static or mobile) or Main Sensor (MS), depending upon the particular scenario.
      \item After aggregation phase, the transmission and remaining energy of BR is calculated. Remaining energy is given as:\\
    Remaining energy = Initial energy - Transmission energy
  \item After all the sensors of first patient are completely checked or if a BR of first patient dies i.e. it consumes its entire energy, then the protocol advances towards checking the next patient.
\end{itemize}

\subsection{Scenarios}
All the five scenarios are shown in Fig. 4. Sink's different positions in scenarios 2 and 5 are shown in TABLE II.
\begin{table}[htbp]
\centering
\begin{tabular}{|c|c|c|}
  \hline
  \textbf{Sink}&\multicolumn{2}{c}{\textbf{Positions}}
  \vline\\
  \cline{2-3}
  &$x$ $(ft)$&$y$ $(ft)$\\
  \hline
  $Sink1$ & $0$ & $10$ \\ \hline
     $Sink2$ & $20$ & $20$  \\ \hline
     $Sink3$ & $40$ & $10$  \\ \hline
     $Sink4$ & $20$ & $0$  \\ \hline
\end{tabular}
\caption{Location of multiple sinks in the hospital ward}
\end{table}

\begin{figure}[htbp]
        \centering
        \begin{subfigure}[b]{0.4\textwidth}
                \centering
                \includegraphics[width=\textwidth]{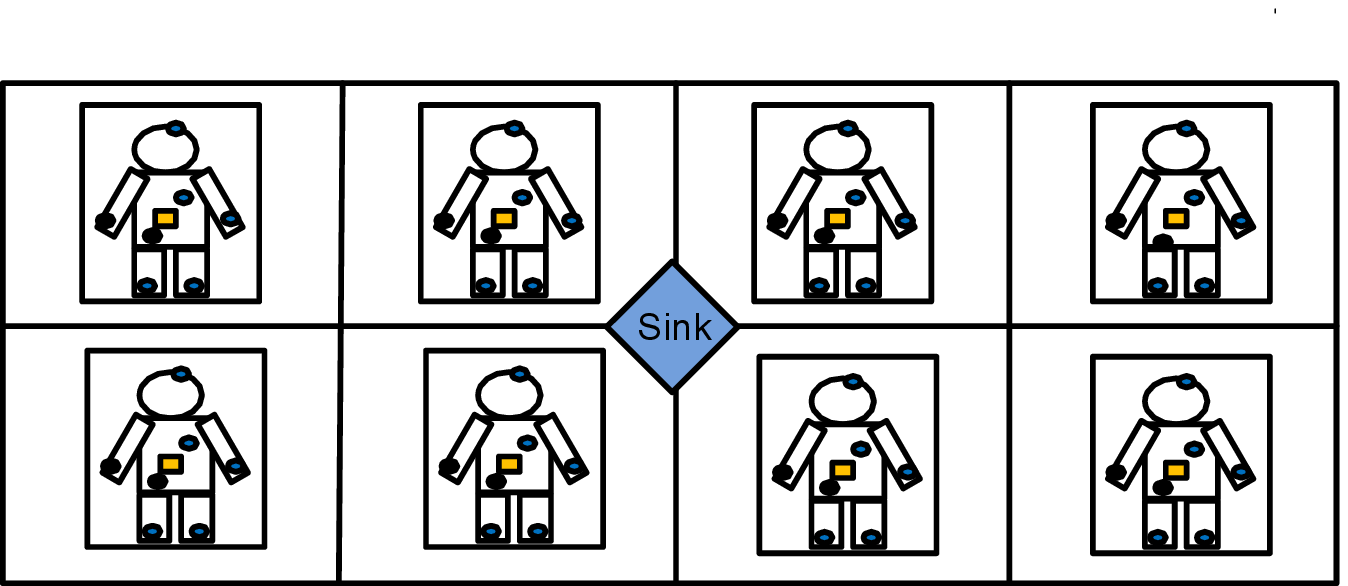}
                \caption{Scenario-1}
                \label{1}
        \end{subfigure}%

        \begin{subfigure}[b]{0.4\textwidth}
                \centering
                 \includegraphics[width=\textwidth]{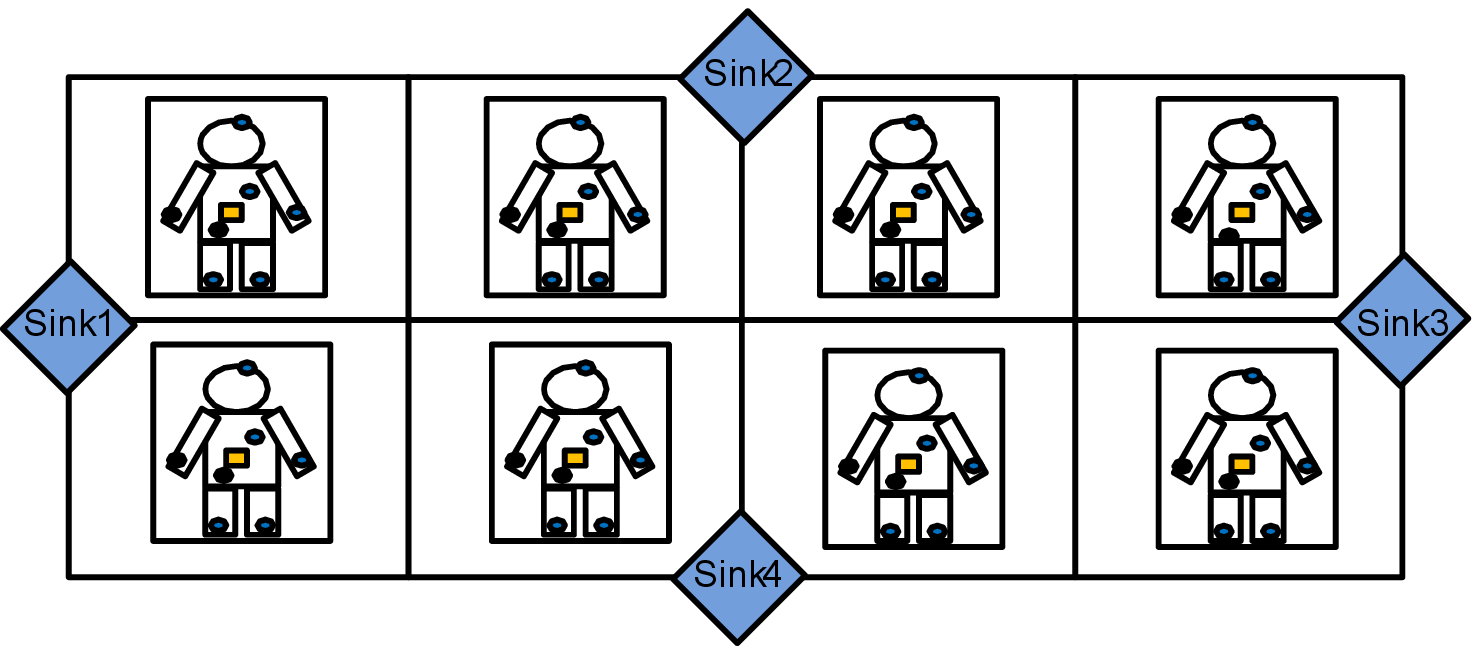}
                \caption{Scenario-2}
                \label{2}
        \end{subfigure}

        \begin{subfigure}[b]{0.4\textwidth}
                \centering
                \includegraphics[width=\textwidth]{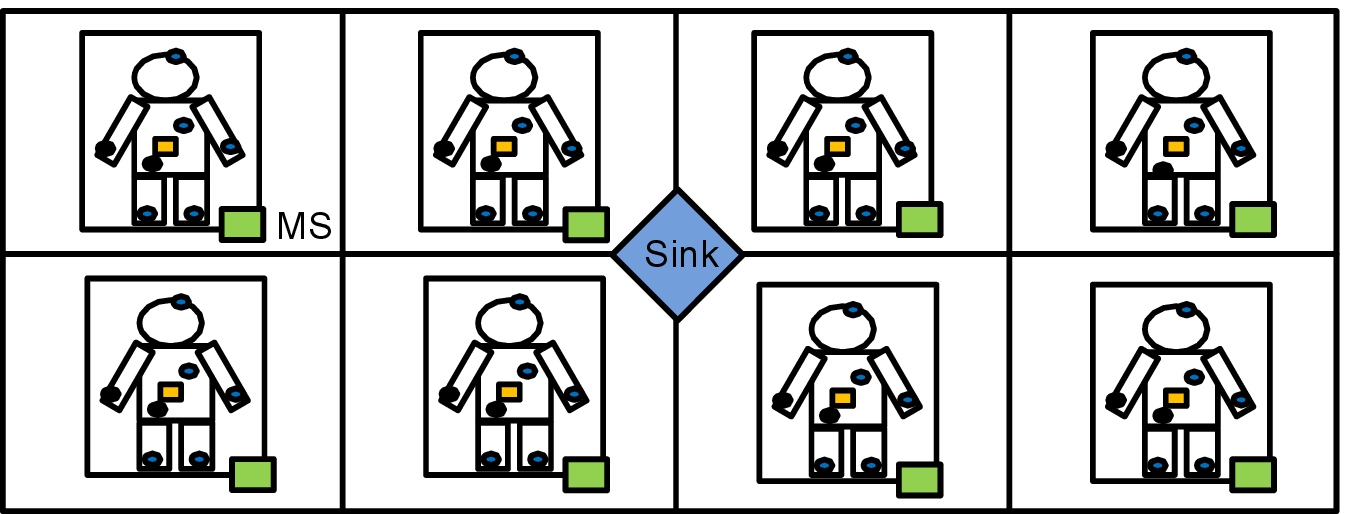}
                \caption{Scenario-3}
                \label{3}
        \end{subfigure}

        \begin{subfigure}[b]{0.4\textwidth}
                \centering
                \includegraphics[width=\textwidth]{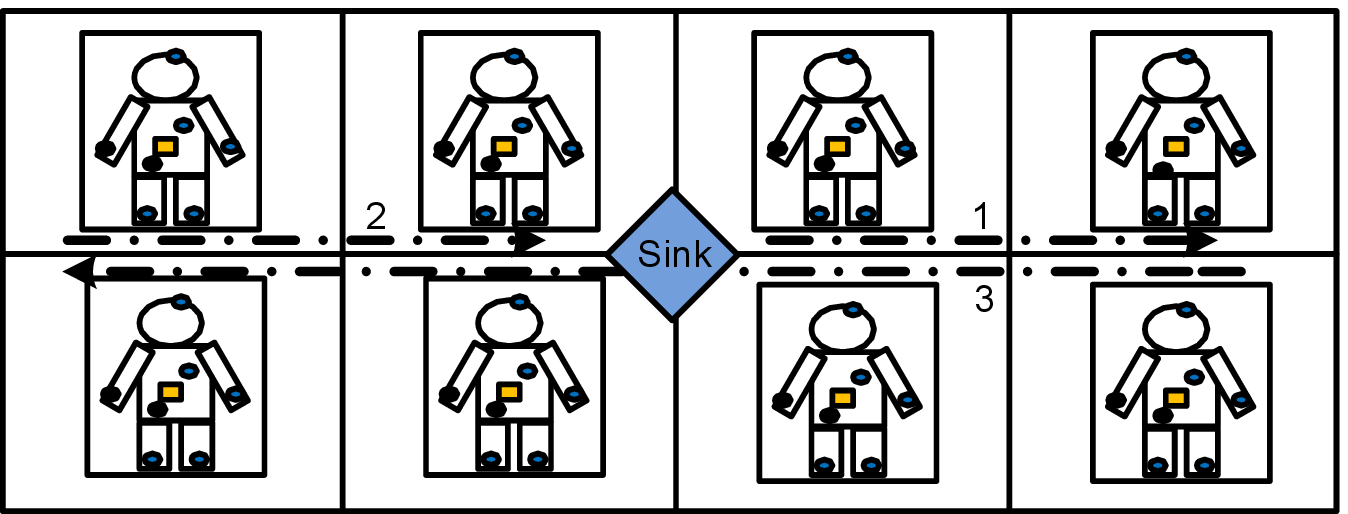}
                \caption{Scenario-4}
                \label{4}
        \end{subfigure}

        \begin{subfigure}[b]{0.4\textwidth}
                \centering
                \includegraphics[width=\textwidth]{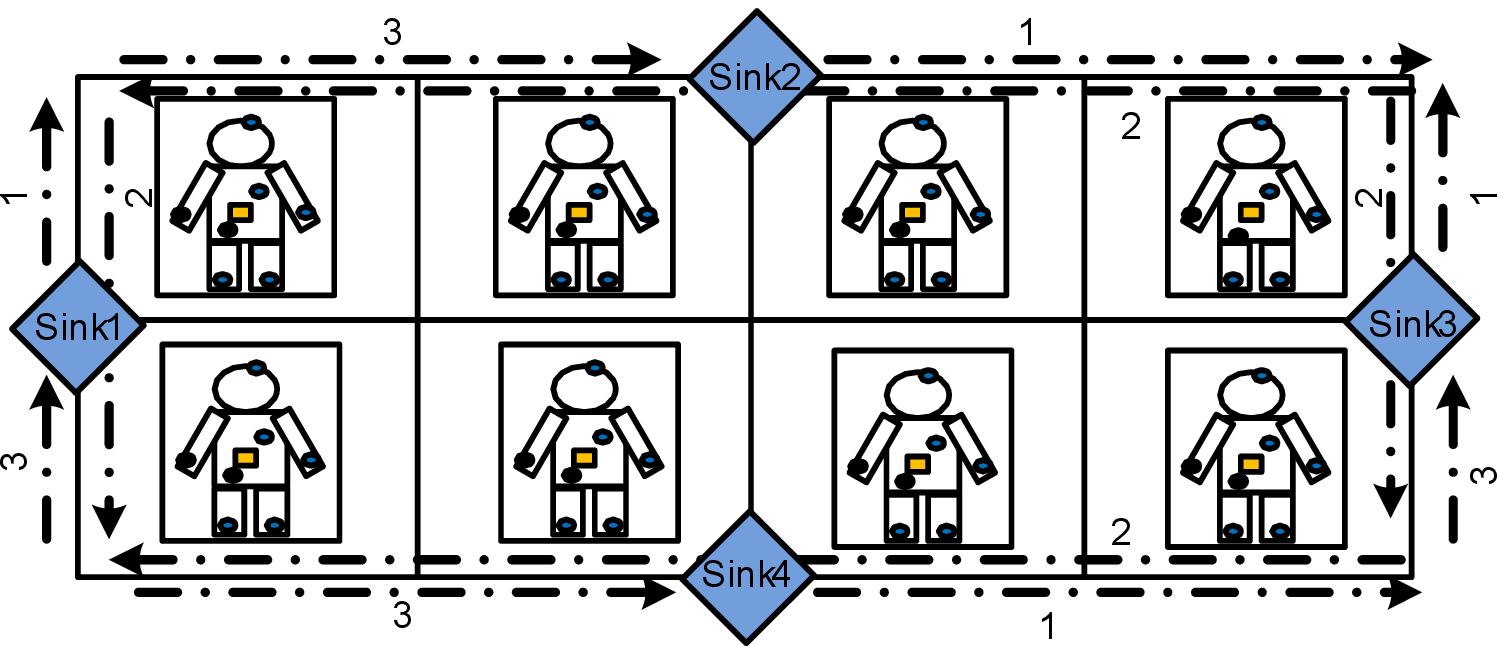}
                \caption{Scenario-5}
                \label{5}
        \end{subfigure}
        \caption{(a): The BSs on each patient carry information and transmit to their respective BR which, then aggregates and relay the received data to the sink located at the center of the ward. The communication flow is from BSs to BR to Sink. (b): Four sinks have been used that are separately deployed in the middle of the walls of the ward. The BSs of each patient, on sensing the vital sign transmit data to their respective BR. The BR checks for the nearest sink by calculating it's distance with each sink. Whichever, sink is found nearest, the BR communicates with that particular sink. The communication flow is from BSs to BR to nearest Sink (Sink1 or Sink2 or Sink3 or Sink4). (c): MS is incorporated on each bed which, can be a PDA type device. The deployment of MS helps the BR to consume little energy as, BR transmits data over shorter distance. However, this scenario increases the delay in the network, as the data traverses through a long route towards the destination node, the Sink. Communication flow is from BSs to BR to MS to Sink. (d): It follows the same communication flow as (a) however, now the sink is made mobile which, moves along the center of ward. (e): Multiple sinks in (b) move around the walls of the ward altogether. In this scenario also, each BR measures it's distance with each sink. Whosoever is found close, the BR starts communicating with that sink. The communication flow is from BSs to BR to the nearest moving Sink (Sink1 or Sink2 or Sink3 or Sink4).}\label{a}
\end{figure}

\subsection{Radio Model and Equations}
The basic radio model proposed for BAN developed in \cite{braem2007need}, is given below:

Equation for transmission energy is given as:
\begin{equation}
  E_{tx}(k,d) =  E_{TXelec}\times k + E_{amp}(n)\times k \times d^n
    \end{equation}
Equation for reception energy is given as:
  \begin{equation}
 E_{rx}(k) =  E_{RXelec} \times k
\end{equation}

where, $E_{tx}$ represents the transmission energy, $E_{rx}$ represents the receiver energy, $E_{TXelec}$ and $E_{RXelec}$ represent the energy which, the radio dissipates to run the circuitry for transmitter and receiver, respectively. $E_{amp}$ represents the energy for transmit amplifier and k is the number of transmitted bits. The values for these parameters are given in TABLE I.

\section{Experimental Results and Discussions}
Comparison plots are taken between the proposed protocol, DARE and the compared protocol, M-ATTEMPT. Different parameters are investigated including remaining alive nodes in the network, residual energy of the network, number of packets sent, received and dropped etc. Both DARE and M-ATTEMPT show the similarity in deploying 7 sensors on each patient. The difference lies in using the different sensor types namely, Body Relays, Main Sensors and Sink in DARE while M-ATTEMPT utilizes only Sink. They are responsible in performing data aggregation and finally sending the data towards the destination node (external network). The protocol runs for 5000 rounds. Graphs for all the parameters are discussed in the following subsections.

\subsection{Alive nodes}
In this subsection, separate graphs are shown for the remaining alive monitoring sensors of both protocols and alive BSs + BRs for DARE along with alive Sensors for M-ATTEMPT.
\begin{itemize}
  \item \emph{Alive - BSs:} The graph for remaining alive BSs for all the scenarios of DARE shows the same response since, all the body sensors transmit data to their respective body relays in the same manner. So, in Fig. 5 we show only one scenario's remaining alive nodes (BSs) and compare with the remaining alive sensor nodes of M-ATTEMPT protocol.
      \begin{figure}[h]
      \begin{center}
      \fbox{
      \includegraphics[height=6cm,width=8cm]{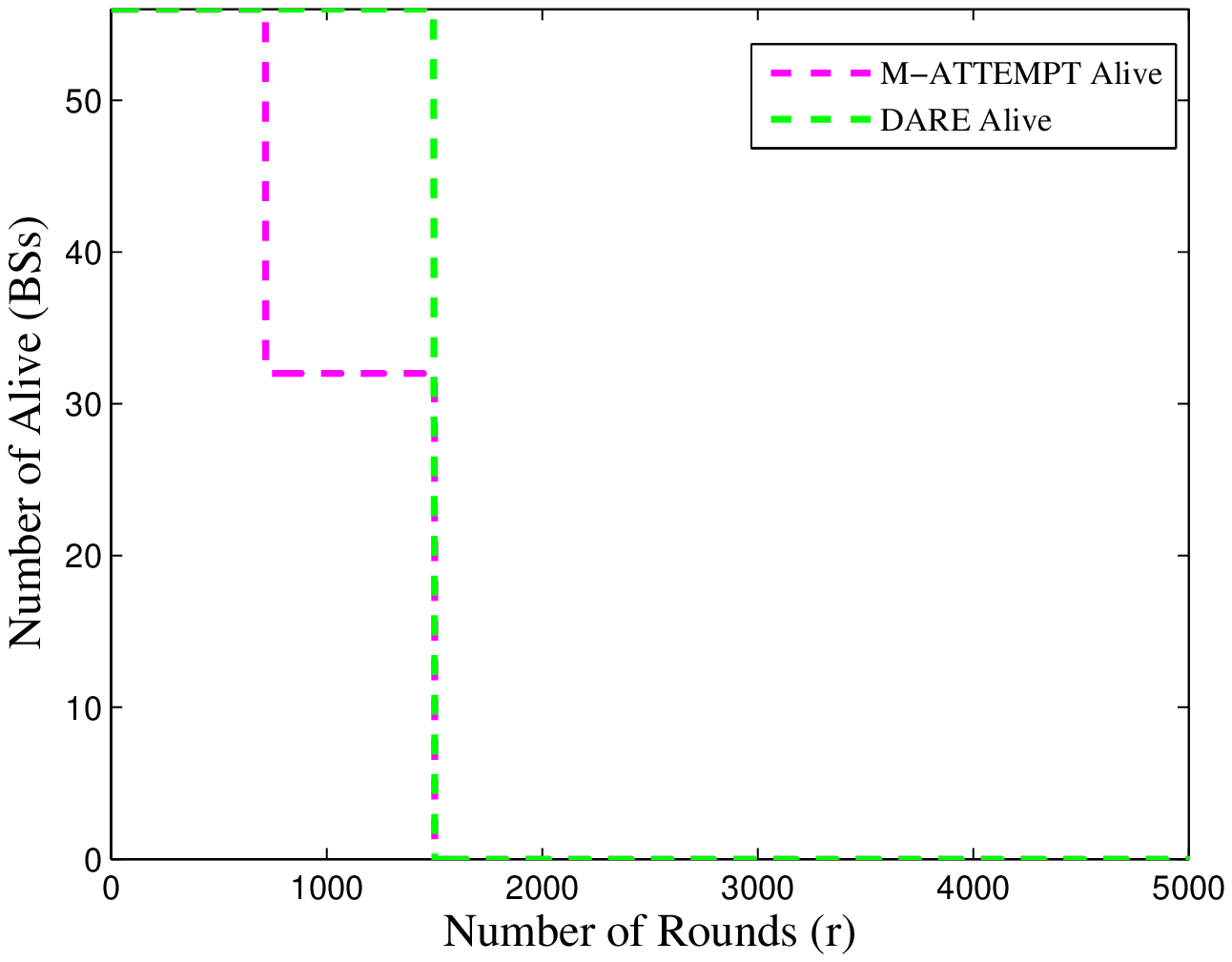}
      }
      \end{center}
      \begin{center}
    \caption{Number of remaining alive nodes (BSs) in the network}
    \end{center}
    \end{figure}

     The time from the establishment of network till the death of first node is known as the stability period. Time is taken in rounds where, each round refers to the time interval in which the protocol operation is performed once.
     In this graph, it is clear that all the 56 nodes in M-ATTEMPT remain alive for only 714 rounds. While, in case of DARE, the nodes remain alive for about 1500 rounds after which they start depleting. In DARE, as soon as the first node dies, the other nodes also die very quickly, thereby, preserving uniformity in the protocol. This means that the energy and the load is uniformly distributed among all the nodes. Hence, DARE has greater stability period and also shows an enhanced network lifetime since, the nodes remain alive for greater number of rounds. The percentage of sensing nodes (BSs in case of DARE and Sensors in case of M-ATTEMPT) to stay alive in terms of rounds increases from “\SI{14}{$\%$}” (M-ATTEMPT) to about “\SI{30}{$\%$}” (DARE). This shows that the stability period of DARE has been increased by “\SI{15}{$\%$}” in relative to M-ATTEMPT.

  \item \emph{Alive - BSs + BRs:} Fig. 6 shows the comparison graph for all scenarios of DARE and M-ATTEMPT. Here, we considered all BSs and BRs (total 64 nodes) of DARE and 56 nodes of M-ATTEMPT.
   \begin{figure}[h]
      \begin{center}
      \fbox{
     \includegraphics[height=6cm,width=8cm]{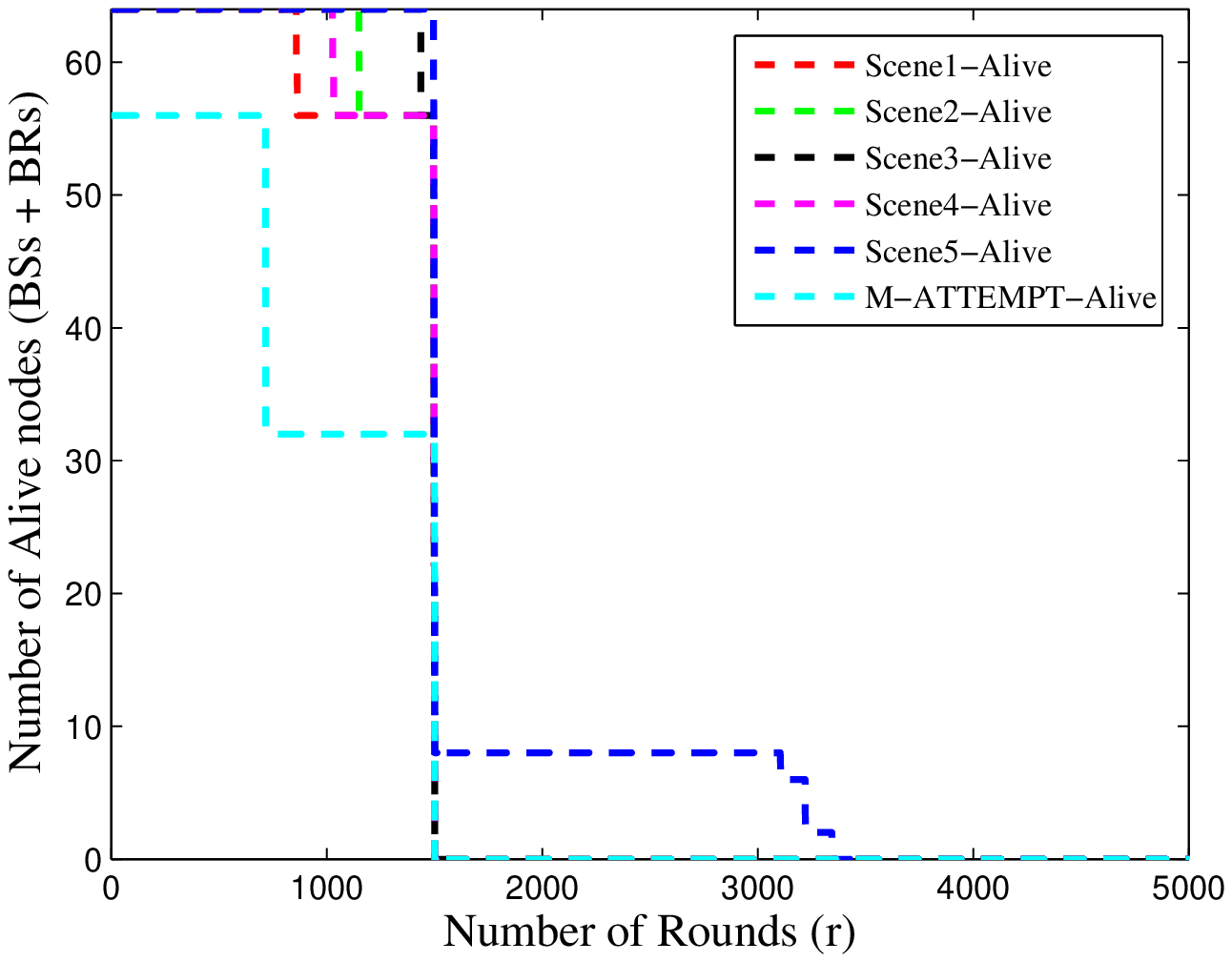}
      }
      \end{center}
      \begin{center}
    \caption{Number of remaining alive nodes (BSs + BRs) in the network}
    \end{center}
    \end{figure}

    In scenario-5, when multiple moving sinks are deployed, the nodes (both BSs and BRs) continue to survive for increased number of rounds i.e. for about 3300 rounds. This is because, multiple sinks at different locations facilitate the BRs to transmit data at shorter distance due to which, the BRs stay alive for some additional rounds due to which they become able to accept more data from the BSs. Of all the scenarios, scenario-5 shows the greatest stability period. Also, a PDA type device which, in this case is the Main Sensor (MS) in scenario-3 attached with the bed, enables the nodes to die with decreased ratio as compared to the other scenarios. In scenario-1, as the sink is static, all the nodes remain alive till only 858 rounds, after which they continue to die. This scenario possesses the least stability period in relative to the other scenarios. Hence, it is the deployment of BRs that help the BSs to continue sending their data. Accordingly, the network lifetime significantly increases.

    In terms of dead nodes, the nodes in case of all scenarios, start to die at later rounds as compared to M-ATTEMPT. It is clear from the graph that mobility of multiple sinks makes possible for the nodes to die later as is, in the case of scenario-5.
\end{itemize}
The period when, the nodes start to decay is termed as, the unstable period of the network. From the above discussion, it can be deduced that the proposed protocol, DARE achieves minimum unstable period as compared to M-ATTEMPT for both the above mentioned cases. Also DARE exhibits maximum network lifetime.

\subsection{Residual energy}
For the case of remaining energy of the nodes in network, we are interested in simulating the results for energy consumption of BSs and Sensors in case of DARE and M-ATTEMPT, respectively. The initial energy of Sensors in M-ATTEMPT and BSs in DARE protocol is kept same i.e. ``\SI{0.3}{J}''. From Fig. 7, the nodes in M-ATTEMPT consume energy very fast as compared to the nodes in DARE. There is a sharp fall-off in the M-ATTEMPT’s energy consumption line i.e. the nodes are utilizing energy with much increased proportion. Because of this early energy utilization, the M-ATTEMPT nodes die out earlier. However, DARE achieves minimum energy consumption as it is facilitated by BRs.
\begin{figure}[h]
  \begin{center}
   \fbox{
\includegraphics[height=6cm,width=8cm]{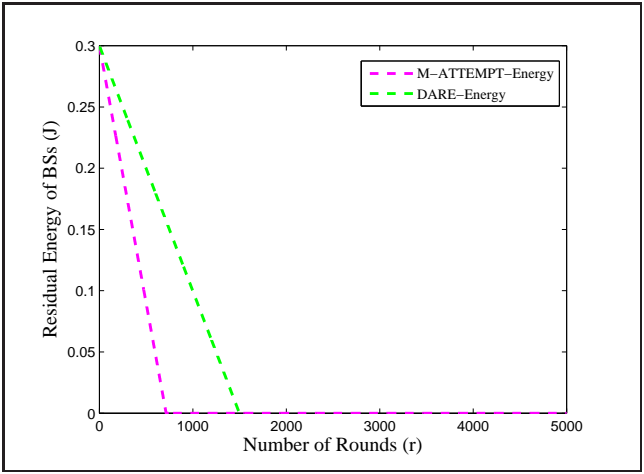}
  }
  \end{center}
  \begin{center}
\caption{Residual energy (BSs) of the network}
\end{center}
\end{figure}

\subsection{Packets sent to sink}
Scenario-5 transmits huge number of packets in the network as compared to the rest scenarios and M-ATTEMPT as shown in Fig. 8. This is due to the reason that the nodes continue to survive for an increased number of rounds. In case of MS attached to the beds and deployment of multiple sinks either static or mobile, BRs are facilitated to cover short distances and consume small amount of energy, thereby, enabling the network to receive significant amount of packets.
\begin{figure}[htbp]
  \begin{center}
   \fbox{
  \includegraphics[height=6cm,width=8cm]{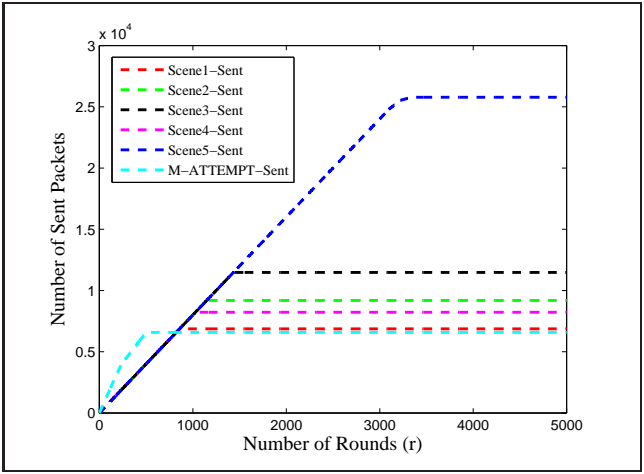}
  }
  \end{center}
  \begin{center}
\caption{Number of packets sent to sink}
\end{center}
\end{figure}

As, the nodes in M-ATTEMPT die out early, minimum number of packets are transmitted in the network.

\subsection{Throughput $(\%)$}
The percentage of total packets received successfully, is known as throughput of the network or packet delivery ratio.
It is expressed as:

\begin{equation}
\text{Throughput(\%) }=  \frac{\text{Number  of  packets  received}}{\text{number  of  packets  sent}}*\text{100}
\end{equation}

In Fig. 9, scenario-5 shows the maximum packet delivery ratio about “\SI{91}{$\%$}” since, it sends greater number of packets in the network in relative to all the other scenarios as well as the M-ATTEMPT protocol. M-ATTEMPT shows the minimum throughput of about “\SI{45}{$\%$}”.

\begin{figure}
  \begin{center}
     \fbox{
  \includegraphics[height=6cm,width=8cm]{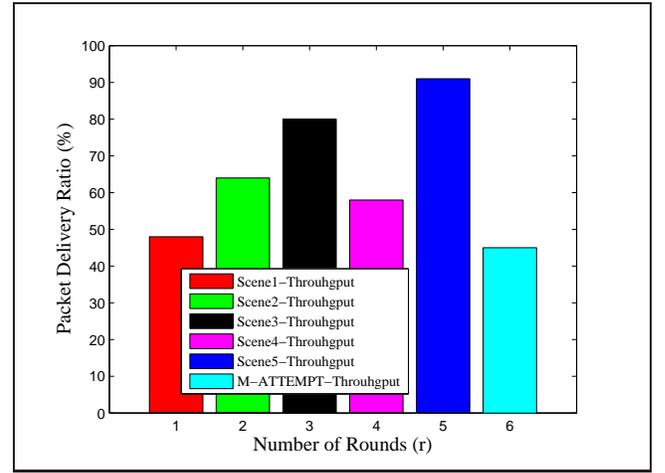}
  }
  \end{center}
  \begin{center}
\caption{Packet delivery ratio}
\end{center}
\end{figure}

\section{DARE and M-ATTEMPT}

TABLE III shows the comparison between overall scenarios of DARE and M-ATTEMPT, with respect to various network parameters.

\begin{table}[h]
\centering
\setlength{\tabcolsep}{12pt}

   \begin{tabular}{| c | c | c | c |}
     \hline
     \textbf{Parameter} & \textbf{DARE} & \textbf{M-ATTEMPT}  \\ \hline
     Stability period& high & low  \\ \hline
     Network lifetime & high & low  \\ \hline
     Energy consumption & minimum & maximum  \\ \hline
     Throughput & high & low  \\ \hline
     Propagation delay & high & low \\ \hline
   \end{tabular}
   \caption{Comparison results between DARE and M-ATTEMPT}
 \end{table}

\section{Conclusion and Future Work}
In this paper, a relaying energy-efficient protocol for heterogeneous networks for monitoring patients is proposed. Some of the sensors monitor data continuously while, others monitor only when a certain threshold level is reached. The protocol defines minimum energy parameters for the sensors to avoid damage to the body tissues. The results clearly show that, the network lifetime and the stability period in terms of more nodes to stay alive (for additional number of rounds) and in terms of reduced energy consumption, our proposed protocol is better than the compared protocol, M-ATTEMPT. The percentage of the number of nodes staying alive, increases from $23$$\%$ (M-ATTEMPT) to about $72$$\%$ (DARE). Also, the proposed protocol provides better throughput in relative to the compared protocol. So, DARE protocol shows great potential in the cases where, human intervention is required to be avoided and also where, huge data transmissions are required. However, M-ATTEMPT provides reduced delay in transmitting packets to the network towards the destination which, makes it a feasible protocol for the networks where there is no room for huge delay.

Future work focuses on estimating the delay in propagating data from body sensors to the destination node in DARE protocol and also to investigate mobility in patients body. Authors in [20-23], work on routing metrics based on ETX (Expected Transmission Count) which, shows better performance than minimum hop count metric, under the availability of link. In view of this, the plan focus to work on routing link metrics as well.

\ifCLASSOPTIONcaptionsoff
  \newpage
\fi

\end{document}